%% file: 0_main.tex
\documentclass[conference, 10pt]{IEEEtran}
\IEEEoverridecommandlockouts
\usepackage{cite}
\usepackage{amsmath,amssymb,amsfonts}
\usepackage{algorithm, algpseudocode}
\usepackage[T1]{fontenc}
\usepackage[utf8]{inputenc}
\usepackage{graphicx}
\usepackage{textcomp}
\usepackage{xcolor}
\usepackage{soul}
\usepackage{algorithm}
\usepackage{algpseudocode}
\usepackage{tablefootnote}
\usepackage{multirow}
\usepackage{url}
\usepackage[normalem]{ulem}
\useunder{\uline}{\ul}{}
\def\BibTeX{{\rm B\kern-.05em{\sc i\kern-.025em b}\kern-.08em
    T\kern-.1667em\lower.7ex\hbox{E}\kern-.125emX}}
\begin{document}

% \title{Presenting Quantum Embedding as a Search Problem and solving it using Genetic Algorithm }
\title{Optimizing Quantum Embedding using Genetic Algorithm for QML Applications}

\author{\IEEEauthorblockN{Koustubh Phalak}
\IEEEauthorblockA{\textit{CSE Department} \\
\textit{Pennsylvania State University}\\
State College, PA \\
krp5448@psu.edu}
\and
\IEEEauthorblockN{Archisman Ghosh}
\IEEEauthorblockA{\textit{CSE Department} \\
\textit{Pennsylvania State University}\\
State College, PA \\
apg6127@psu.edu}
\and
\IEEEauthorblockN{Swaroop Ghosh}
\IEEEauthorblockA{\textit{School of EECS} \\
\textit{Pennsylvania State University}\\
State College, PA \\
szg212@psu.edu}}
% \and
% \and
% \and
% \and
% \and

\maketitle

\begin{abstract}
Quantum Embeddings (QE) is an important component of Quantum Machine Learning (QML) algorithms to load classical data present in Euclidean space onto quantum Hilbert space, which are then later forwarded to the Parametric Quantum Circuit (PQC) for training and finally measured to compute the cost classically. The performance of the QML algorithm can vary according to the type of QE used, and also on mapping of features within the embedding (i.e., onto the qubits). This provides motivation to search for the optimal quantum embedding (i.e., feature to qubit mapping). Typically this problem is presented as an optimization problem, where the quantum embedding has trainable parameters and the optimal embedding is found out by training the embedding. In this work, we present identification of the optimal embedding as a search problem rather than an optimization problem. We show that for fixed number of qubits and model initialization of a Quantum Neural Network (QNN), different mapping of features onto qubits via QE changes the final performance of the QML algorithm. We propose Genetic Algorithm (GA) based search to find the optimal mapping of the features to the qubits. We perform experiments to find the optimal QE for binary classes of MNIST and Tiny Imagenet datasets and compare the results with randomly selected feature to qubit mapping (under identical number of runs). Our results show that GA-based approach performs better than random selection for MNIST (Tiny Imagenet) by 0.33-3.33 (0.5-3.36) higher average fitness score with up to 8.1\%-15\% (for MNIST) and 5.3\%-8.8\% (for Tiny Imagenet) less runtime. For both the datasets increasing the qubit counts marginally affects the GA fitness implying that the GA is scalable both in terms of dataset and in terms of QNN size. Compared to existing methodologies such as Quantum Embedding Kernel (QEK), Quantum Approximate Optimization Algorithm (QAOA)-based embedding and Quantum Random Access Code (QRAC), GA performs better by 1.003X,  1.03X and 1.06X respectively.
\end{abstract}


\input{1_Introduction.tex}

\input{2_Background.tex}

\input{3_Optimality.tex}
\input{4_Idea.tex}
\input{6_Conclusion.tex}

% \section*{Acknowledgment}

% The preferred spelling of the word ``acknowledgment'' in America is without 
% an ``e'' after the ``g''. Avoid the stilted expression ``one of us (R. B. 
% G.) thanks $\ldots$''. Instead, try ``R. B. G. thanks$\ldots$''. Put sponsor 
% acknowledgments in the unnumbered footnote on the first page.

\bibliographystyle{unsrt}
\bibliography{refs}

\end{document}

%% file: 1_Introduction.tex
\section{Introduction} 
%\hl{todo: edit text, add diagrams in intro, add citations, discuss}
Quantum Machine Learning (QML) combines quantum computing and machine learning to enhance data processing capabilities, especially for complex datasets that challenge classical algorithms. In the noisy intermediate-scale quantum (NISQ) era of quantum devices \cite{Preskill_2018}, with limited qubit count and coherence times, implementing efficient QML algorithms is a significant challenge. A key aspect of QML algorithms is the embedding of classical data into quantum states. Quantum embeddings transform data points into high-dimensional quantum feature spaces, enabling effective data representation in the quantum Hilbert Space. However, finding an optimal embedding pattern is essential due to the limitations inherent in NISQ devices. Efficient embeddings allow features with similar structural information to be placed across qubits in a way that increases the performance of the QML model on NISQ-era hardware. %This paper explores a genetic algorithm-based approach to discover optimal quantum embedding patterns, tailored for specific datasets and NISQ constraints, to maximize the performance of QML applications.

% \hl{pull fig. 4 up in intro as fig. 1 and explain the problem simple mindedly. Check abstract as well. Mention that we considered angle encoding since its the most popular and it may work for basis but not amplitude. Mention this in limitation subsection as well.}
\begin{figure}[t]
    \centering
    \includegraphics[width=\linewidth]{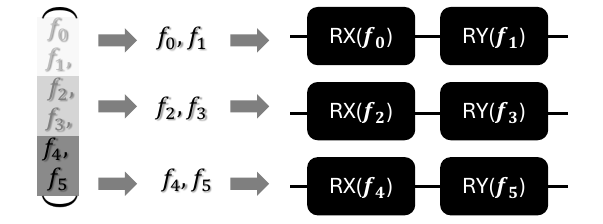}
    %\vspace{-10pt}
    \caption{Example showing how permutation ($f_0$, $f_1$, $f_2$, $f_3$, $f_4$, $f_5$) is embedded on 3 qubits using angle embedding. A pair of two features is mapped on each qubit, such that the first feature is embedded using an $RX$ gate, and the second feature is embedded using an $RY$ gate.}
    \label{fig:embedding_example}
    % \vspace{-10pt}
\end{figure}
\subsection{What is Quantum Embedding?}
\begin{figure}[t]
    \centering
    \includegraphics[width=0.8\linewidth]{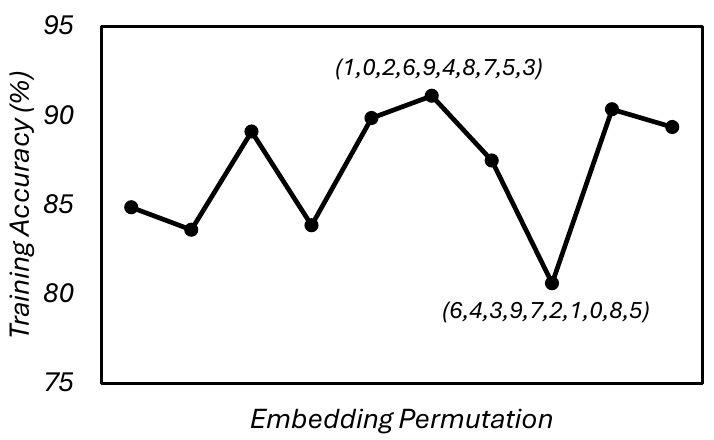}
    %\vspace{-10pt}
    \caption{Plot representing the variation of training accuracy with respect to different feature embedding patterns during training of class labels 2 and 6 from the MNIST dataset on a 5-qubit classifier. The data labels represent the sequential ordering of the features (numbered 0-9) on 5 qubits (features 1 and 0 on physical qubit 1; 2 and 6 on physical qubit 2, and so on).}
    \label{fig:var}
    \vspace{-10pt}
\end{figure}

Quantum Embedding (QE) is a specific circuit involving rotation gates to embed the classical data in the Hilbert space. From a quantum hardware perspective, it is the mapping of the logical quantum states that embed the data onto the physical qubits of the quantum computing architecture. Different embedding techniques (angle, basis, and amplitude) impact the embedding and performance of the QML models differently. \textbf{Angle embedding} encodes classical data in the parameters of single-qubit rotations $RX(\theta)$, $RY(\theta)$, and $RZ(\theta)$, providing a straightforward approach but can become complex for higher-dimensional data, as each feature requires a separate rotation gate, potentially increasing circuit depth. \textbf{Basis embedding} assigns each data point to a specific quantum basis state. For binary data, basis embedding is natural, mapping directly to computational basis states. However, its applicability is limited in higher-dimensional settings, as it requires a high number of qubits to represent each unique data point, making it less scalable for larger datasets. \textbf{Amplitude embedding} is more compact and powerful for complex datasets, encoding data as the amplitudes of a quantum state. This approach provides a dense representation, embedding multiple data features into a single quantum state, but it requires normalized data and can be challenging to implement due to its high circuit complexity. In our paper, we experiment with angle embedding as it is the most popular embedding technique owing to the ease of implementation and data representation. From Fig. \ref{fig:embedding_example}, we observe a very simple example of angle embedding. A data point having six features ($f_0$-$f_5$) is embedded into 3 physical qubits using $RX$ and $RY$ gates. It is important to note that the pattern of embedding the features may vary and would provide varying results (Fig. \ref{fig:var}). In this example, the total possible number of embedding permutations would be $6!$, making it essential to optimize the data embedding to maximize the performance of the QML circuit.

\subsection{Motivation}
In QML models, not only the choice of the embedding procedure but also the pattern of embedding dictates the performance of the circuit. Embedding related features of a data point mapped on various qubits affects the training accuracy of the QML model differently making it important to search for an optimal embedding pattern. This is evident from Fig. \ref{fig:var}. We train a 5-qubit classifier (details in Section V) on the MNIST dataset (labels 2 and 6) and show the variation in training accuracy with different embedding patterns. We encode two features on each qubit (using $RX(\theta)$ and $RY(\theta)$ gates), totaling $10!$ possible embedding permutations from which we plot 10 random samples to show the variation. Even after training under identical circumstances, the training accuracy varies between 80\% and 92\% depending on the logical to physical mapping of the features. Therefore, it is significant to search for an optimal pattern for embedding the classical data in the quantum states of the QML circuit.

Framing quantum embedding selection as a \textit{search problem} rather than a traditional optimization problem provides unique advantages. Optimization methods typically explore a continuous parameter space for defining quantum embedding kernels that embed the data and often require gradient-based techniques to converge to optimal kernel configurations. However, by treating this task as a search problem, we can utilize several heuristic search methods like genetic algorithms to efficiently explore all discrete possibilities of determining an optimal embedding structure. A search procedure is less resource-intensive than training a quantum embedding kernel.

% This search-based approach is especially advantageous for NISQ hardware, as it allows systematic exploration of diverse, hardware-efficient embedding patterns. The search framework supports flexibility, making it possible to prioritize embedding configurations that work effectively within qubit limitations, reduce noise susceptibility, and align with specific data structures. Thus, treating embedding selection as a search problem can yield noise-resilient, hardware-compatible embeddings that optimize QML performance in a targeted, resource-aware manner.%\subsection{Contribution}

\begin{figure}[t]
    \centering
    \includegraphics[width=1\linewidth]{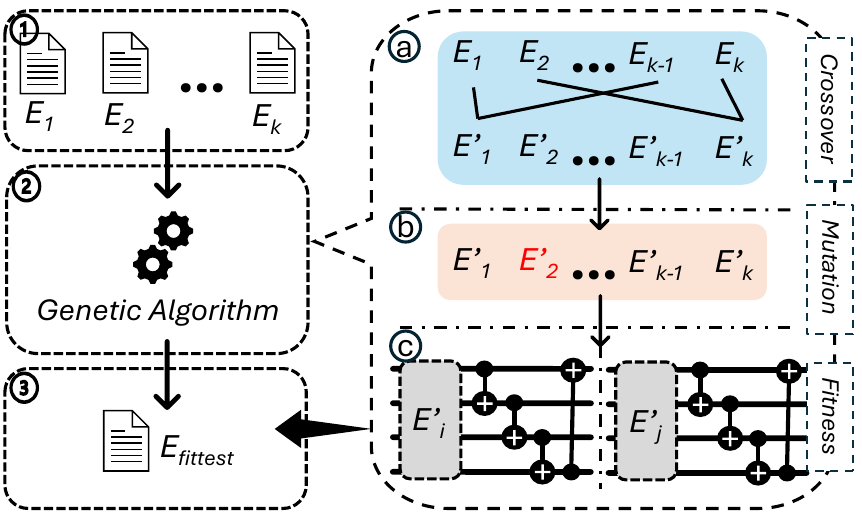}
    %\vspace{-10pt}
    \caption{A demonstration of the workflow of our proposed idea. In the figure, (1) shows the search space of all possible embedding permutations labeled $E_1$ to $E_k$; 2(a) demonstrates the procedure of crossover. For example, $E_1$ and $E_{k-1}$ are crossed to obtain the \textit{offspring} $E'_1$, and so on; 2(b) shows the idea of mutation where two features are selected and swapped at random in the embedding ($E'_2$ in the example); 2(c) shows the state where the fitness score of all individuals in the population is compared by preparing the embedding and simulating the QML circuit, and the fittest embedding permutation ($E_{fittest}$) is output in (3).}
    \label{fig:flow}
    \vspace{-10pt}
\end{figure}

\emph{To the best of our knowledge, this is the first attempt at presenting the embedding of classical data in the Hilbert Space as a search problem.} We develop a genetic algorithm-based approach to select the embedding that performs best in terms of training and test accuracy. We observe the workflow in Fig. \ref{fig:flow}. $E_1$-$E_k$ represents the search space for the optimal embedding permutation. We run the genetic algorithm on this search space with the fitness function involving the training and test accuracies of the QML networks formed with the embedding permutations after performing crossover and mutation on them, to obtain an optimal solution ($E_{fittest}$) to the search problem. Furthermore, we test our proposed algorithm by implementing multi-qubit classifiers trained on MNIST and Tiny Imagenet datasets and compare with randomly selected logical feature to qubit mapping. Additionally, we compare our proposed approach with existing works such as \cite{schuld2021supervised,lloyd2020quantumembeddingsmachinelearning,qrac} for MNIST dataset.

\subsection{Paper Structure}

Section II provides a background on quantum computing and the genetic algorithm. Section III presents the idea of quantum embedding selection as a search problem and shows that there exists an optimal solution to this problem.
Section IV details the proposed algorithm followed by the experimental results. Section V concludes the paper.

%% file: 2_Background.tex
\section{Background}
\subsection{Quantum Computing}
A qubit is analogous to a classical bit and is the fundamental unit of quantum computing. However, a qubit can be in the superposition of the two states—0 or 1, unlike a classical bit, which exists in one state at a time. In Dirac notation, the state of a qubit is denoted as $|\psi\rangle = \alpha|0\rangle + \beta|1\rangle$, where $\alpha$ and $\beta$ are complex numbers that must satisfy the normalization condition $|\alpha|^2 + |\beta|^2 = 1$. The computational basis states are represented as $ |0\rangle = [1 \ 0]^T$ and $|1\rangle = [0 \ 1]^T$. When dealing with multiple qubits, $n$ qubits can represent a quantum state within a $2^n$-dimensional space, with basis states ranging from $|0\dots0\rangle$ to $|1\dots1\rangle$. An $n$-qubit quantum state can be expressed as $|\psi_n\rangle = \sum_{i=0}^{2^n-1} a_i |i\rangle$, where the coefficients $a_i$ satisfy the normalization condition $\sum_{i=0}^{2^n-1} |a_i|^2 = 1$.

\begin{figure}[t]
    \centering
    \includegraphics[width=1\linewidth]{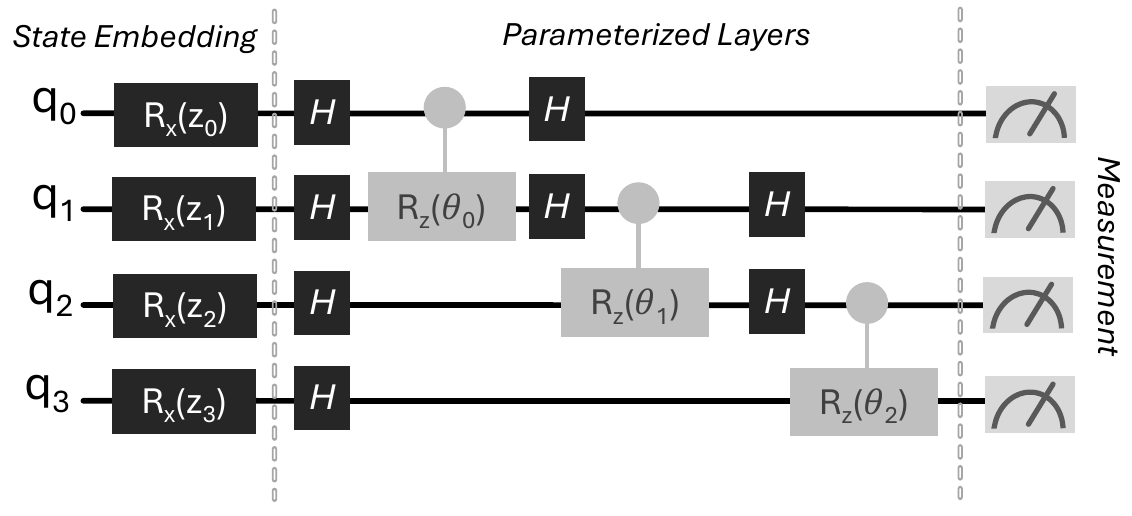}
    %\vspace{-10pt}
    \caption{The figure represents a circuit diagram of a QNN. In state embedding, the $RX(z_i)$ gates are used for basis encoding to map the classical data onto the Hilbert Space. The parameterized layers comprise a series of $CRZ(\theta_i)$ gates that provide the entanglement as well as a finer-grained search into the Hilbert Space. The workflow of the QNN is concluded with a measurement operator on all the qubits to derive an output.}
    \label{fig:pqc}
    \vspace{-10pt}
\end{figure}

\subsection{Quantum Neural Networks}
Quantum Neural Networks (QNNs) integrate quantum computing with machine learning \cite{Schuld_2015}, enabling tasks like regression and classification by encoding classical data into qubit states. QNN design (Fig. \ref{fig:pqc}) involves several steps: \textbf{Quantum Embedding} encodes classical data into the Hilbert space using quantum states, with techniques such as amplitude encoding, which normalizes data into qubit amplitudes; angle encoding, which converts data into rotation angles for qubits; and basis encoding, mapping binary data to computational basis states. \textbf{Parameterized Quantum Circuits (PQCs)} form the core of the QNN, featuring tunable quantum gates, including single-qubit rotations $RX(\theta)$, $RY(\theta)$, and $RZ(\theta)$, and multi-qubit rotations like $CRX(\phi)$, $CRY(\phi)$, and $CRZ(\phi)$ with parameters controlling qubit interactions. Entanglement between qubits boosts computational capabilities, allowing PQCs to perform complex transformations similar to layers in classical neural networks. \textbf{Measurement} extracts classical information by collapsing quantum states to final qubit states (0 or 1), where the probabilities of each basis state are measured to derive outputs, processed classically. The QNN workflow starts with preprocessing and encoding data into quantum states, continues with processing in the PQC, and optimizes PQC parameters iteratively using classical algorithms by minimizing a loss function until the QNN converges.

\begin{figure*}[t]
    \centering
    \includegraphics[width=\linewidth]{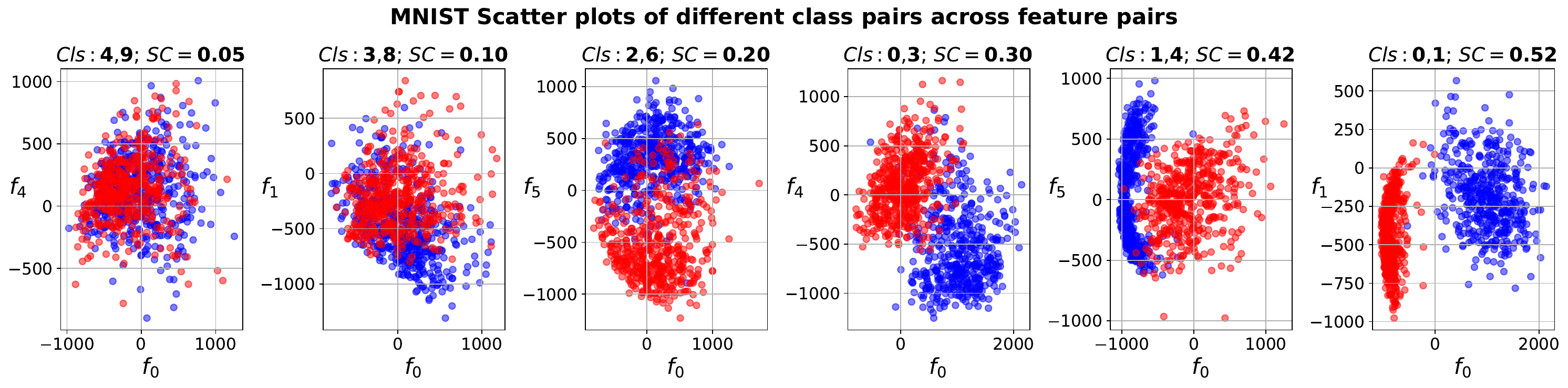}
    \vspace{-3mm}
    \caption{Various MNIST binary classes with varying SC values. The SC value increases from leftmost subplot ($0.05$) to the rightmost subplot ($0.52$).}
    \label{fig:mnist_varying_sc}
    \vspace{-4mm}
\end{figure*}

\begin{figure*}[t]
    \centering
    \includegraphics[width=\linewidth]{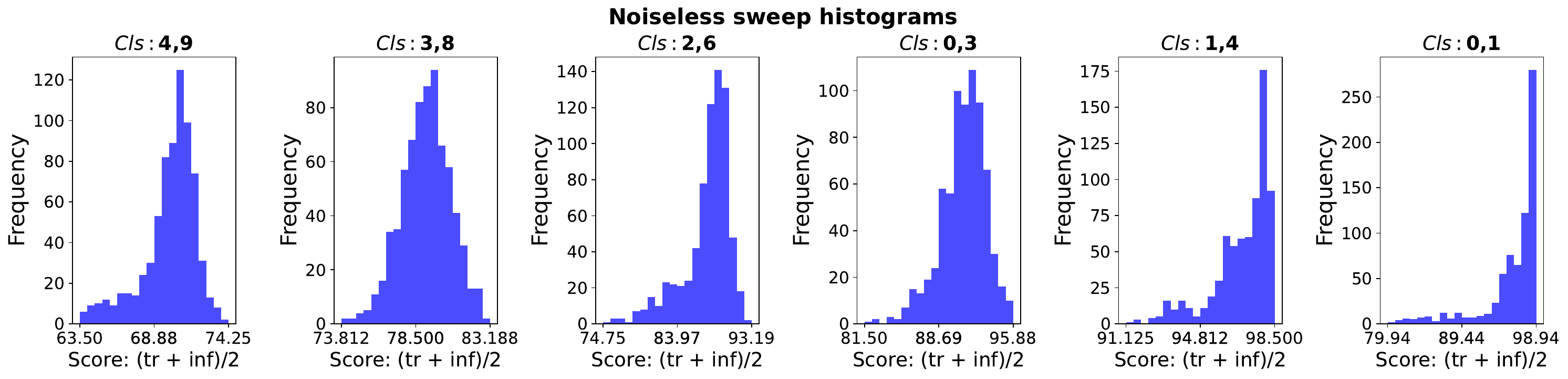}
    \vspace{-3mm}
    \caption{Histograms of sweeping all permutations of embeddings for various binary classes of MNIST dataset under noiseless conditions. The mean of histogram increases with SC value for binary class. Note, Cls: x, y denotes classes x and y.}
    \label{fig:mnist_brute_force_sweep_noiseless}
    \vspace{-4mm}
\end{figure*}

\begin{figure*}[t]
    \centering
    \includegraphics[width=\linewidth]{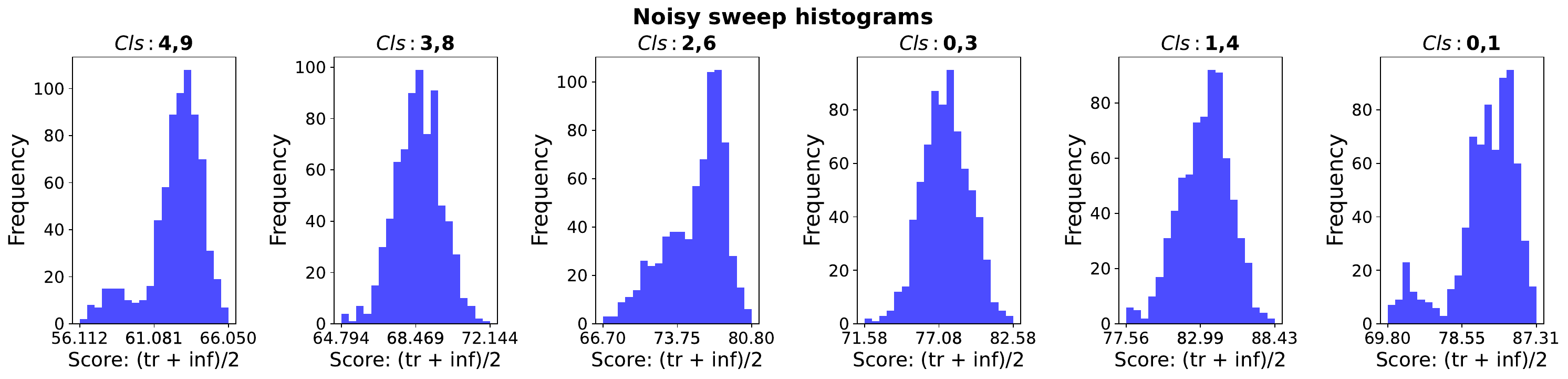}
    \vspace{-3mm}
    \caption{Histograms of sweeping all permutations of embeddings for various binary classes of MNIST dataset under noisy conditions. Compared to noiseless conditions, the peformance of every embedding reduces. Once again, the mean of histogram increases with SC value. Note, Cls: x, y denotes classes x and y.}
    \label{fig:mnist_brute_force_sweep_noisy}
    \vspace{-4mm}
\end{figure*}

\subsection{Genetic Algorithm}
A genetic algorithm (GA) is a heuristic search algorithm inspired by natural selection \cite{forrest1996genetic}, used to solve optimization and search problems by mimicking the process of biological evolution. In a genetic algorithm, a population of candidate solutions, or \textit{individuals}, is evolved over successive generations to find near-optimal solutions to a problem. Each \textit{individual} in the population represents a potential solution corresponding to specific features or decision variables. The process begins with an initial population, which can be randomly generated or based on prior knowledge. The quality of each \textit{individual} is evaluated against a predefined objective function, termed \textit{fitness}. The GA then applies selection, crossover, and mutation operators to evolve the population. Selection favors individuals with higher \textit{fitness} scores, allowing them to contribute their \textit{genes} to the next generation. Crossover combines the attributes of two selected \textit{individuals} of the previous generation to create new offspring, while mutation introduces small, random changes to some \textit{individuals} to ensure diversity in the population. As a search algorithm, GAs are particularly tailored for navigating large, complex, and discrete search spaces, making them great candidates for determining the optimal embedding patterns in QML circuits for efficient training and inferencing on NISQ devices.

\subsection{Related Work}
All existing techniques for determining efficient embedding patterns of classical data on the quantum encoding circuit have been extensively studied as an optimization problem. Quantum classification by training the quantum feature maps (embeddings) rather than focusing on post-embedding measurements \cite{lloyd2020quantumembeddingsmachinelearning} is one such idea. This method, termed quantum metric learning, aims to separate data classes maximally within a Hilbert space, using optimal measurements based on distance metrics (e.g., Helstrøm for trace distance, overlap for Hilbert-Schmidt distance), minimizing resource use on NISQ devices, facilitating efficient classification by simplifying the measurement stage in quantum machine learning workflows. Another take on trainable embeddings is to use Quantum Random Access Codes (QRACs) \cite{qrac_orig}. A ($n,m$) QRAC works by encoding $n$ classical bits into $m$ qubits, allowing any single bit from the original set to be retrieved with a probability $p>1/2$. Integrating QRACs with quantum metric learning to train the quantum embedding enhances feature embeddings by adapting QRAC to support classification tasks with complex Boolean functions, which traditional QRAC struggles with \cite{qrac}. The authors demonstrate improved classification accuracy in variational quantum classifiers (VQCs) for datasets with discrete features, achieving robust performance on real-world datasets. 
The concept of trainable quantum embedding kernels (QEKs) is also introduced to enhance data classification in quantum machine learning \cite{qek}. Using kernel-target alignment and noise mitigation, it optimizes quantum embeddings on NISQ devices, demonstrating improved robustness and accuracy in embedding performance through experiments on real quantum hardware. This has prompted research in meta-heuristic algorithms to develop optional quantum kernels based on combinatorial optimization techniques \cite{incudini2023automaticeffectivediscoveryquantum}, improving accuracy over manual kernel design, especially in anomaly detection for high-energy physics applications. Aligning the trainable quantum kernels by sub-sampling methods was an attempt to reduce the computational costs by training on smaller subsets of data at each iteration \cite{Sahin_2024}. Using quantum models solely as kernel methods to make data linearly separable and using Support Vector Machine (SVM) to classify this data was proposed in \cite{schuld2021supervised}.
Despite providing improved results, the existing methods rely on complex gradient calculation for the optimization procedure which is often computationally expensive. Employing heuristic search algorithms like GA sidesteps the computational overhead and provides an optimal search pattern tailored for implementing QML circuits in NISQ devices.

%% file: 3_Optimality.tex
\section{Presence of Optimality for QE Search}

 In this section, we empirically show the existance of an optimality while mapping input data features to qubits during embedding. 
 \subsection{Basic setup}
 For this search problem, we restrict ourselves to angle embedding, where we embed two features on each qubit, one using RX gate and second using RY gate. Additionally, we use Strongly Entangling Layer (SEL) \cite{schuld2020circuit} for the PQC and perform expectation value measurements in the Pauli-Z basis. Consider the case of $3$ qubits, which would imply embedding a total of $6$ features. We refer to these features as $f_0$, $f_1$, $f_2$, $f_3$, $f_4$ and $f_5$ respectively. Considering all permutations, we get a total of $6! = 720$ ways to embed features onto these $3$ qubits. We represent an embedding permutation as a tuple of features and for example, we show a visualization of how the permutation ($f_0$, $f_1$, $f_2$, $f_3$, $f_4$, $f_5$) is embedded in Fig. \ref{fig:embedding_example}. We perform sweep on all $720$ possible permutations with fixed input data (fixed train-test split always), fixed model initialization and fixed hyperparameters ($5$ epochs, $3\times10^{-3}$ learning rate) where in every permutation of the sweep, we train the model and note the final training and inferencing performance. For dataset, we use different binary classes of MNIST dataset with varying Silhouette Coefficients (SC) \cite{kaufman2009finding}. The system used for running these experiments has the following configuration: CPU Intel i9-13900K, GPU Nvidia RTX 3090 24 GB, and RAM 64 GB. Note, SC is a value ranging in $[-1,+1]$, where $-1$ value represents perfect overlap of all class clusters while $+1$ value represents non-overlapping class clusters. In general, higher the SC value, the easier it becomes for model to classify the data. We show scatter plots of different binary classes of MNIST dataset (reduced from $28 \times 28$ to $6$ features using PCA \cite{pearson1901liii}) with varying SC values in Fig. \ref{fig:mnist_varying_sc}. As we move from left (classes $4$, $9$) to right (classes $0$, $1$), the SC increases due to better separation among the classes. For every pair of classes, we perform the aforementioned sweep under noiseless conditions (Fig. \ref{fig:mnist_brute_force_sweep_noiseless}). For each embedding permutation, we note a combined score which is the mean of training and inferencing accuracies. This is done to make sure that both training and inferencing accuracies are high.

 \begin{algorithm}[t]
\caption{Genetic Algorithm for optimal QE search}
\label{alg:genetic_algo}
\begin{algorithmic}[1]
\Require Population size $s_{pop} = 20$, generations $g = 5$, crossover rate $cr = 0.8$, retention rate $rr = 0.1$, mutation rate $mr = 0.001$
\State $features = (0,1,2,3,4,5)$
\State $pop = [perm(features)$ for $i$ in $range(s_{pop})]$ $\#$init. pop
\State $fs = [fitness(f)$ for $f$ in $pop]$ $\#$fitness scores
\For{each generation $i = 1$ to $g$}
    \State $sort(fs, reverse=True)$ $\#$ fs descending
    \State $os = []$ $\#$ offspring
    \State \raisebox{.5pt}{\textcircled{\raisebox{-.9pt} {\textbf{1}}}} $os.add(pop[:rr \times s_{pop}]$) $\#$ retain top $rr \times s_{pop}$
    \State \raisebox{.5pt}{\textcircled{\raisebox{-.9pt} {\textbf{2}}}} $parents = select(pop, fs)$ $\#$ select parents
    \For{$j$ in $[rr\times s_{pop} : len(pop)]$} $\#$ remaining length
        \State $p_1$, $p_2$ $= random(parents, 2)$
        \If{$random() < cr$}
            \State \raisebox{.5pt}{\textcircled{\raisebox{-.9pt} {\textbf{3}}}} $child$ $=$ $cross(p_1, p_2)$ $\#$ apply crossover
        \Else
            \State $child = p_1$ $\#$ set child as copy of one parent
        \EndIf
        \If{$random() < mr$}
            \State \raisebox{.5pt}{\textcircled{\raisebox{-.9pt} {\textbf{4}}}} $child$ $=$ $mutate(child)$ $\#$ apply mutation
        \EndIf
        \State $os.append(child)$ $\#$ add child to offspring
    \EndFor
    \State $pop = os$ $\#$ replace population 
\EndFor
\State \textbf{return} best individual from $pop$ based on fitness
\end{algorithmic}
\end{algorithm}

\subsection{Analysis}
 From the results of the sweep under noiseless conditions, we observe that some embedding configurations perform better than others. For example, for binary classes 2 and 6 the embedding ($f_5$, $f_2$, $f_0$, $f_3$, $f_1$, $f_4$) gives best combined score of $93.18$ (training $93.37\%$, inferencing $93.0\%$), while embedding ($f_2$, $f_5$, $f_3$, $f_1$, $f_0$, $f_4$) performs the worst with combined score of $74.75$ (training $72\%$, inferencing $77.5\%$). Such a considerable varation of $\sim20$ in performance under extremely controlled environment (fixed model parameters, fixed dataset, fixed hyperparameters, only embedding changes) proves that there exists an optimal embedding configuration that can yield the best possible performance. The existence of optimality is an important finding because when running these QNN models on NISQ computers which are plagued by noise, this variation may fluctuate owing to degraded performance under noise. We show this aspect by performing simulations under noisy conditions. For each embedding permutation, we take average training and inferencing accuracies of $10$ runs of training while computing combined score in order to account for extreme fluctuations caused by noise \footnote{Henceforth, for all noisy simulations we take mean training and inferencing accuracies of $10$ runs.}. We select noise model of \textit{FakeBrisbane()} backend from IBM quantum with coupling configuration $0-1-2$ as target coupling \footnote{We similarly select \textit{FakeBrisbane()} for all future noisy simulations with same and higher qubit counts.}. From the histograms of noisy sweep plots (Fig. \ref{fig:mnist_brute_force_sweep_noisy}), we immediately note degraded performance compared to noiseless conditions for all embedding permutations. Once again considering the case of binary classes 2 and 6, we note best performing embedding ($f_4$, $f_1$, $f_2$, $f_3$, $f_5$, $f_0$) performs the best with combined score $80.8$ (training $79.3\%$, inferencing $82.3\%$), while embedding ($f_3$, $f_1$, $f_2$, $f_5$, $f_4$, $f_0$) performs the worst with combined score $66.7$ (training $65.65\%$, inferencing $67.75$). Overall, we see a fluctuation of $\sim14$ in the combined score, which is slightly less compared to noiseless conditions yet indicative of effect of noise on the overall performance variation.

%We now proceed to finding the optimal configuration. To do so, we propose using a GA that finds a configuration that is either close to best-performing configuration or even better than that. 

%% file: 4_Idea.tex
\section{Proposed Idea and Results}

\subsection{GA based search}
Since sweep based search for optimal solution for the QE problem is computationally intractable, we use GA. First, we describe the selection, crossover and mutation steps of the GA, followed by the entire algorithm.

\begin{table*}[t]
\centering
\caption{MNIST low qubit count results. Here, (c$_1$,c$_2$,Q) = (class$\_$1, class$\_$2, $\#$qubits), rs $=$ random selection, ga $=$ Genetic Algorithm.}
\label{tab:mnist_low_qubit_count}
\begin{tabular}{|l|llllll|llllll|}
\hline
          & \multicolumn{6}{c|}{\textbf{Noiseless fitness scores}}                                                                                                                                                                                                                                  & \multicolumn{6}{c|}{\textbf{Noisy fitness scores}}                                                                                                                                                                                                                                      \\ \hline
\multicolumn{1}{|c|}{\multirow{2}{*}{\textbf{\begin{tabular}[c]{@{}c@{}}(c$_1$,c$_2$,\\ Q)\end{tabular}}}} & \multicolumn{3}{c|}{\textbf{RS}}                                                                                & \multicolumn{2}{c|}{\textbf{GA}}                                           & \multirow{2}{*}{\textbf{\begin{tabular}[c]{@{}l@{}}Improvement\\ (GA - RS)\end{tabular}}} & \multicolumn{3}{c|}{\textbf{RS}}                                                                                & \multicolumn{2}{c|}{\textbf{GA}}                                           & \multirow{2}{*}{\textbf{\begin{tabular}[c]{@{}l@{}}Improvement\\ (GA - RS)\end{tabular}}} \\ \cline{2-6} \cline{8-12}
\multicolumn{1}{|c|}{}                                                                                     & \multicolumn{1}{c|}{\textbf{mean}} & \multicolumn{1}{c|}{\textbf{best}} & \multicolumn{1}{l|}{\textbf{runtime}} & \multicolumn{1}{c|}{\textbf{best}} & \multicolumn{1}{l|}{\textbf{runtime}} &                                                                                          & \multicolumn{1}{c|}{\textbf{mean}} & \multicolumn{1}{c|}{\textbf{best}} & \multicolumn{1}{l|}{\textbf{runtime}} & \multicolumn{1}{c|}{\textbf{best}} & \multicolumn{1}{l|}{\textbf{runtime}} &                                                                                          \\ \hline
\textit{(4,9,3)}                                                                                           & \multicolumn{1}{l|}{70.04}         & \multicolumn{1}{l|}{73.43}         & \multicolumn{1}{l|}{6.9 min}                 & \multicolumn{1}{l|}{72.12}         & \multicolumn{1}{l|}{7 min}                 & -1.3                                                                                     & \multicolumn{1}{l|}{64.25}         & \multicolumn{1}{l|}{67.15}         & \multicolumn{1}{l|}{2.32 hrs}                 & \multicolumn{1}{l|}{65.50}         & \multicolumn{1}{l|}{2.32 hrs}                 & -1.7                                                                                     \\ \hline
\textit{(3,8,3)}                                                                                           & \multicolumn{1}{l|}{79.30}         & \multicolumn{1}{l|}{82.06}         & \multicolumn{1}{l|}{6.7 min}                 & \multicolumn{1}{l|}{82.37}         & \multicolumn{1}{l|}{6.8 min}                 & 0.3                                                                                      & \multicolumn{1}{l|}{70.44}         & \multicolumn{1}{l|}{72.78}         & \multicolumn{1}{l|}{2.32 hrs}                 & \multicolumn{1}{l|}{70.88}         & \multicolumn{1}{l|}{2.3 hrs}                 & -1.9                                                                                     \\ \hline
\textit{(2,6,3)}                                                                                           & \multicolumn{1}{l|}{87.49}         & \multicolumn{1}{l|}{91.93}         & \multicolumn{1}{l|}{6.9 min}                 & \multicolumn{1}{l|}{91.18}         & \multicolumn{1}{l|}{6.4 min}                 & -0.8                                                                                     & \multicolumn{1}{l|}{77.59}         & \multicolumn{1}{l|}{81.70}         & \multicolumn{1}{l|}{2.3 hrs}                 & \multicolumn{1}{l|}{78.60}         & \multicolumn{1}{l|}{2.3 hrs}                 & -3.1                                                                                     \\ \hline
\textit{(0,3,3)}                                                                                           & \multicolumn{1}{l|}{90.88}         & \multicolumn{1}{l|}{95.5}          & \multicolumn{1}{l|}{7.1 min}                 & \multicolumn{1}{l|}{93.68}         & \multicolumn{1}{l|}{6.3 min}                 & -1.8                                                                                     & \multicolumn{1}{l|}{79.94}         & \multicolumn{1}{l|}{83.60}         & \multicolumn{1}{l|}{2.32 hrs}                 & \multicolumn{1}{l|}{80.46}         & \multicolumn{1}{l|}{2.26 hrs}                 & -3.1                                                                                     \\ \hline
\textit{(1,4,3)}                                                                                           & \multicolumn{1}{l|}{96.90}         & \multicolumn{1}{l|}{98.31}         & \multicolumn{1}{l|}{6.8 min}                 & \multicolumn{1}{l|}{98.25}         & \multicolumn{1}{l|}{7.2 min}                 & -0.1                                                                                     & \multicolumn{1}{l|}{84.43}         & \multicolumn{1}{l|}{88.62}         & \multicolumn{1}{l|}{2.38 hrs}                 & \multicolumn{1}{l|}{86.79}         & \multicolumn{1}{l|}{2.34 hrs}                 & -1.8                                                                                     \\ \hline
\textit{(0,1,3)}                                                                                           & \multicolumn{1}{l|}{95.64}         & \multicolumn{1}{l|}{98.62}         & \multicolumn{1}{l|}{7 min}                 & \multicolumn{1}{l|}{98.68}         & \multicolumn{1}{l|}{6.9 min}                 & 0.1                                                                                      & \multicolumn{1}{l|}{32.12}         & \multicolumn{1}{l|}{88.56}         & \multicolumn{1}{l|}{2.37 hrs}                 & \multicolumn{1}{l|}{86.28}         & \multicolumn{1}{l|}{2.28 hrs}                 & -2.3                                                                                     \\ \hline
\textit{(4,9,4)}                                                                                           & \multicolumn{1}{l|}{73.81}         & \multicolumn{1}{l|}{80.62}         & \multicolumn{1}{l|}{8.9 min}                 & \multicolumn{1}{l|}{79.06}         & \multicolumn{1}{l|}{8.2 min}                 & -1.6                                                                                     & \multicolumn{1}{l|}{66.36}         & \multicolumn{1}{l|}{69.03}         & \multicolumn{1}{l|}{3.23 hrs}                 & \multicolumn{1}{l|}{71.63}         & \multicolumn{1}{l|}{2.78 hrs}                 & 2.6                                                                                      \\ \hline
\textit{(3,8,4)}                                                                                           & \multicolumn{1}{l|}{83.74}         & \multicolumn{1}{l|}{87.25}         & \multicolumn{1}{l|}{8.6 min}                 & \multicolumn{1}{l|}{87.5}          & \multicolumn{1}{l|}{8 min}                 & 0.3                                                                                      & \multicolumn{1}{l|}{73.73}         & \multicolumn{1}{l|}{76.70}         & \multicolumn{1}{l|}{3.22 hrs}                 & \multicolumn{1}{l|}{78.12}         & \multicolumn{1}{l|}{2.78 hrs}                 & 1.4                                                                                      \\ \hline
\textit{(2,6,4)}                                                                                           & \multicolumn{1}{l|}{87.07}         & \multicolumn{1}{l|}{91.31}         & \multicolumn{1}{l|}{8.6 min}                 & \multicolumn{1}{l|}{89.68}         & \multicolumn{1}{l|}{7.8 min}                 & -1.6                                                                                     & \multicolumn{1}{l|}{76.42}         & \multicolumn{1}{l|}{80.50}         & \multicolumn{1}{l|}{3.22 hrs}                 & \multicolumn{1}{l|}{82.13}         & \multicolumn{1}{l|}{2.82 hrs}                 & 1.6                                                                                        \\ \hline
\textit{(0,3,4)}                                                                                           & \multicolumn{1}{l|}{90.21}         & \multicolumn{1}{l|}{95.56}         & \multicolumn{1}{l|}{8.9 min}                 & \multicolumn{1}{l|}{93.06}         & \multicolumn{1}{l|}{7.9 min}                 & -2.5                                                                                     & \multicolumn{1}{l|}{77.86}         & \multicolumn{1}{l|}{81.59}         & \multicolumn{1}{l|}{3.16 hrs}                 & \multicolumn{1}{l|}{83.93}         & \multicolumn{1}{l|}{2.76 hrs}                 & 2.3                                                                                      \\ \hline
\textit{(1,4,4)}                                                                                           & \multicolumn{1}{l|}{97.04}         & \multicolumn{1}{l|}{98.75}         & \multicolumn{1}{l|}{8.8 min}                 & \multicolumn{1}{l|}{98.62}         & \multicolumn{1}{l|}{8 min}                 & -0.1                                                                                     & \multicolumn{1}{l|}{82.77}         & \multicolumn{1}{l|}{87.59}         & \multicolumn{1}{l|}{3.21 hrs}                 & \multicolumn{1}{l|}{90.08}         & \multicolumn{1}{l|}{2.83 hrs}                 & 2.5                                                                                      \\ \hline
\textit{(0,1,4)}                                                                                           & \multicolumn{1}{l|}{95.14}         & \multicolumn{1}{l|}{98.56}         & \multicolumn{1}{l|}{8.9 min}                 & \multicolumn{1}{l|}{98.55}         & \multicolumn{1}{l|}{7.9 min}                 & 0                                                                                        & \multicolumn{1}{l|}{80.94}         & \multicolumn{1}{l|}{85.75}         & \multicolumn{1}{l|}{3.13 hrs}                 & \multicolumn{1}{l|}{90.61}         & \multicolumn{1}{l|}{2.86 hrs}                 & 4.9                                                                                      \\ \hline
\textit{(4,9,5)}                                                                                           & \multicolumn{1}{l|}{74.38}         & \multicolumn{1}{l|}{80.93}         & \multicolumn{1}{l|}{11 min}                 & \multicolumn{1}{l|}{78.87}         & \multicolumn{1}{l|}{9.9 min}                 & -2.1                                                                                     & \multicolumn{1}{l|}{65.42}         & \multicolumn{1}{l|}{68.68}         & \multicolumn{1}{l|}{4.06 hrs}                 & \multicolumn{1}{l|}{70.35}         & \multicolumn{1}{l|}{3.72 hrs}                 & 1.7                                                                                      \\ \hline
\textit{(3,8,5)}                                                                                           & \multicolumn{1}{l|}{82.48}         & \multicolumn{1}{l|}{86.87}         & \multicolumn{1}{l|}{10.7 min}                 & \multicolumn{1}{l|}{85.12}         & \multicolumn{1}{l|}{9.9 min}                 & -1.8                                                                                       & \multicolumn{1}{l|}{69.98}         & \multicolumn{1}{l|}{73.6}          & \multicolumn{1}{l|}{4.09 hrs}                 & \multicolumn{1}{l|}{76.03}         & \multicolumn{1}{l|}{3.58 hrs}                 & 2.4                                                                                      \\ \hline
\textit{(2,6,5)}                                                                                           & \multicolumn{1}{l|}{86.68}         & \multicolumn{1}{l|}{92.56}         & \multicolumn{1}{l|}{10.8 min}                 & \multicolumn{1}{l|}{90.37}         & \multicolumn{1}{l|}{9.7 min}                 & -2.2                                                                                     & \multicolumn{1}{l|}{73.31}         & \multicolumn{1}{l|}{76.73}         & \multicolumn{1}{l|}{4.06 hrs}                 & \multicolumn{1}{l|}{79.65}         & \multicolumn{1}{l|}{3.55 hrs}                 & 2.9                                                                                      \\ \hline
\textit{(0,3,5)}                                                                                           & \multicolumn{1}{l|}{88.69}         & \multicolumn{1}{l|}{95.5}          & \multicolumn{1}{l|}{11.1 min}                 & \multicolumn{1}{l|}{91.68}         & \multicolumn{1}{l|}{9.7 min}                 & -3.8                                                                                       & \multicolumn{1}{l|}{73.38}         & \multicolumn{1}{l|}{78.08}         & \multicolumn{1}{l|}{4.03 hrs}                 & \multicolumn{1}{l|}{81.73}         & \multicolumn{1}{l|}{3.55 hrs}                 & 3.7                                                                                      \\ \hline
\textit{(1,4,5)}                                                                                           & \multicolumn{1}{l|}{95.89}         & \multicolumn{1}{l|}{99.18}         & \multicolumn{1}{l|}{10.9 min}                 & \multicolumn{1}{l|}{98.37}         & \multicolumn{1}{l|}{9.7 min}                 & -0.8                                                                                     & \multicolumn{1}{l|}{79.31}         & \multicolumn{1}{l|}{83.81}         & \multicolumn{1}{l|}{4.04 hrs}                 & \multicolumn{1}{l|}{86.25}         & \multicolumn{1}{l|}{3.55 hrs}                 & 2.4                                                                                      \\ \hline
\textit{(0,1,5)}                                                                                           & \multicolumn{1}{l|}{95.64}         & \multicolumn{1}{l|}{98.68}         & \multicolumn{1}{l|}{11.1 min}                 & \multicolumn{1}{l|}{98.37}         & \multicolumn{1}{l|}{9.7 min}                 & -0.3                                                                                     & \multicolumn{1}{l|}{78.75}         & \multicolumn{1}{l|}{86.33}         & \multicolumn{1}{l|}{4.08 hrs}                 & \multicolumn{1}{l|}{87.0}          & \multicolumn{1}{l|}{3.6 hrs}                 & 0.7                                                                                      \\ \hline
\end{tabular}

% \vspace{0.1cm}
% \hspace{0.5cm} \raggedright \footnotesize{$^3$The brackets show the features encoded in a single qubit. For example, in ($f_4$,$f_0$),($f_1$,$f_5$),($f_2$,$f_3$); 
% ($f_4$,$f_0$) gets mapped to qubit 0, ($f_1$,$f_5$) to qubit 1, and ($f_2$,$f_3$) to qubit 2.}
\end{table*}

\textbf{\textit{Selection:}} Tournament selection is employed to select parents based on fitness scores of the current population. For the GA, we define the fitness score of an embedding permutation to be its combined score, as described earlier. The tournament size is set to 2, so randomly two parents are selected from the population. The parent with higher overall fitness score is selected from the selection process. For example, for binary MNIST classes 2 and 6, embedding permutation ($f_1$, $f_3$, $f_4$, $f_0$, $f_2$, $f_5$) has a fitness score of $82.63$ and embedding permutation ($f_3$, $f_2$, $f_1$, $f_4$, $f_5$, $f_0$) has fitness score of $91.18$. So, if the tournament selection randomly selects these two parents, ($f_3$, $f_2$, $f_1$, $f_4$, $f_5$, $f_0$) will be selected due to its larger fitness score. The selection process is repeated until the number of parents selected for crossover becomes equal to the population size.

\textbf{\textit{Crossover:}} Two parents are used to perform a crossover and create a child. Suppose the two parents are $p_1$ and $p_2$. We randomly select an \textit{individual} feature from $p_1$, and the features of $p_1$ up to and including the \textit{individual} are added to the child. For the leftover features, features of $p_2$ that are not present in a child are selected and added to the child. For example, consider $p_1$ is ($f_3$, $f_2$, $f_1$, $f_4$, $f_5$, $f_0$), $p_2$ is ($f_0$, $f_5$, $f_1$, $f_2$, $f_3$, $f_4$), and randomly selected pivot is $f_2$ in $p_1$. So part of child from $p_1$ is ($f_3$, $f_2$). The leftover features not present in the child are $f_0$, $f_1$, $f_4$, and $f_5$. These features are present in the order ($f_0$, $f_5$, $f_1$, $\_$, $\_$, $f_4$) ($f_2$, $f_3$ excluded for illustrative purposes). Appending these features in this order, the child generated from $p_1$ and $p_2$ becomes ($f_3$, $f_2$, $f_0$, $f_5$, $f_1$, $f_4$).

\textbf{\textit{Mutation:}} We incorporate mutation in the generated child (with low probability). If the mutation happens, two random features are selected and swapped. Consider the child from crossover described previously ($f_3$, $f_2$, $f_0$, $f_5$, $f_1$, $f_4$). If randomly selected features are $f_2$ and $f_4$, then after mutation, the child becomes ($f_3$, $f_4$, $f_0$, $f_5$, $f_1$, $f_2$).

\textbf{\textit{Algorithm:}} We define population size $s_{pop}=20$, generations $g=5$, crossover rate $cr=0.8$, retention rate $rr=0.1$, and mutation rate $mr=0.001$. We start by generating $s_{pop}$ number of random embedding permutations for which fitness scores are computed. Based on these initial fitness scores, the \textit{indiviuduals} of the population are sorted in decreasing order, and the following steps are performed iteratively for $g$ generations: \raisebox{.5pt}{\textcircled{\raisebox{-.9pt} {\textbf{1}}}} top $rr$ fraction of best performing embedding permutations in current population are retained. \raisebox{.5pt}{\textcircled{\raisebox{-.9pt} {\textbf{2}}}} Selection is performed to select parents from the current population. \raisebox{.5pt}{\textcircled{\raisebox{-.9pt} {\textbf{3}}}} For remaining $1-rr$ fraction of population, children are generated from selected parents via crossover. \raisebox{.5pt}{\textcircled{\raisebox{-.9pt} {\textbf{4}}}} Finally, with mutation rate $mr$, children are mutated. We show the overall algorithm in Algorithm \ref{alg:genetic_algo} for input data having $6$ features ($3$ qubits). 

\begin{figure}[t]
    \centering
    \includegraphics[width=\linewidth]{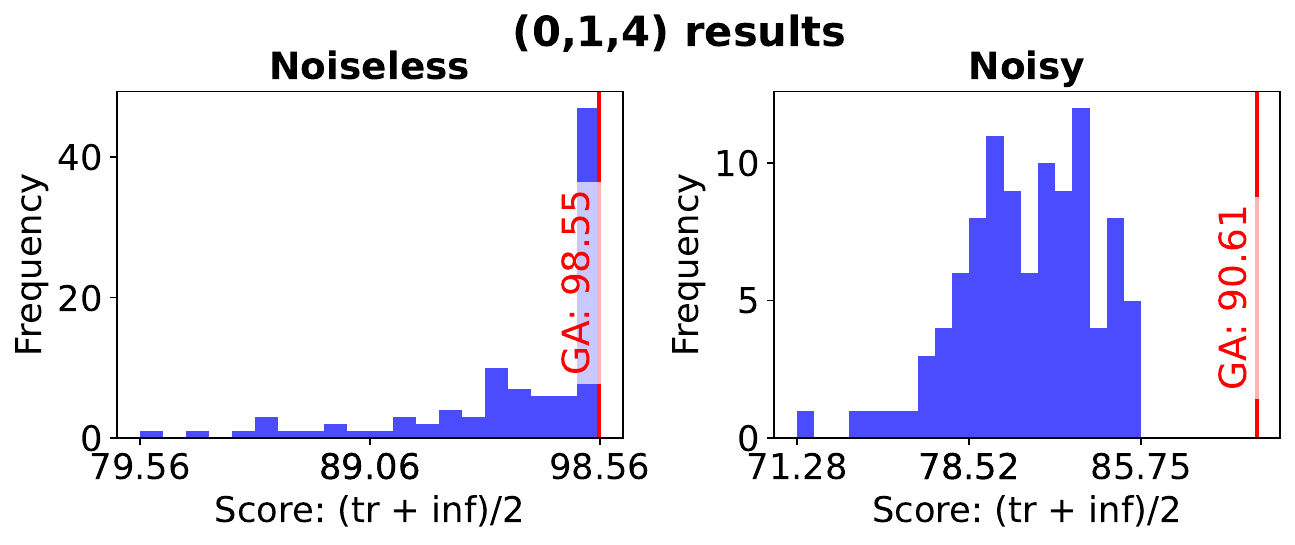}
    \caption{Noiseless and noisy plots for ($0$,$1$,$4$) MNIST case. Optimal fitness score found by GA is highlighted on red line along with value.}
    \vspace{-5mm}
    \label{fig:0_1_4_results}
\end{figure}

\subsection{Setup for results}

With $20$ population size and $5$ generations, the GA trains a total of $20 \times 5=100$ embedding permutations. We compare the performance of GA by randomly selecting $100$ embedding permutations and training them for two datasets: MNIST and Tiny Imagenet \cite{le2015tiny} (reduced to lower dimension using UMAP \cite{leland2018uniform}). Additionally, we breakdown the analysis of results by low qubit count and high qubit count of the QNN. For low qubit count we select $3$, $4$ and $5$ qubits, while for high qubit count we choose $6$, $7$ and $8$ qubits for the QNN. We present results for Tiny Imagenet (noisy only) to show scalability of the algorithm to larger dataset and high qubit count to show scalability in terms of number of qubits. Additionally, we also study two SCs $0.05$ and $0.51$ to examine their effect of GA fitness score.

Overall, we perform four different analyses: \raisebox{.5pt}{\textcircled{\raisebox{-.9pt} {\textbf{1}}}} \textit{low qubit count MNIST}, with binary classes $4-9$, $3-8$, $2-6$, $0-3$, $1-4$ and $0-1$ under both noiseless and noisy conditions; \raisebox{.5pt}{\textcircled{\raisebox{-.9pt} {\textbf{2}}}} \textit{low qubit count Tiny Imagenet} with binary classes bell pepper-orange (SC$=$$0.05$) and sulphur butterfly-alp (SC$=$$0.51$) only noisy conditions; \raisebox{.5pt}{\textcircled{\raisebox{-.9pt} {\textbf{3}}}} \textit{high qubit count MNIST}, with binary classes $4-9$ and $0-1$, only noisy conditions; \raisebox{.5pt}{\textcircled{\raisebox{-.9pt} {\textbf{4}}}} \textit{high qubit count Tiny Imagenet} with binary classes bell pepper-orange and sulphur butterfly-alp, only noisy conditions. Compared to the other three analyses, the analysis performed in low qubit count MNIST case is the most extensive because it is has both less complex dataset as well as less number of qubits, which makes the analysis easier and more fundamental. The analyses for other cases can be extrapolated from this particular case with respect to the trend observed. Note, fitness score difference of seemingly low values (such as 1-3) is significant since it potentially indicates a change in both training as well as inferencing accuracy by a similar amount.

\begin{table}[t]
\centering
\caption{GA results for Tiny Imagenet (low and high qubit count) and MNIST (high qubit count) datasets. BP$=$bell paper, ON$=$orange, SB$=$sulphur butterfly, AP$=$alp, SC$=$silhouette coefficient.}
\label{tab:three_scenario_table}
\begin{tabular}{|ccccccc|}
\hline
\multicolumn{2}{|c|}{\textbf{}}                                                                                     & \multicolumn{5}{c|}{\textbf{Noisy fitness scores}}                                                                                                                                         \\ \hline
\multicolumn{1}{|c|}{\multirow{2}{*}{\textbf{(c$_1$,c$_2$,Q)}}} & \multicolumn{1}{c|}{\multirow{2}{*}{\textbf{SC}}} & \multicolumn{2}{c|}{\textbf{RS}}                                           & \multicolumn{2}{c|}{\textbf{GA}}                                           & \multirow{2}{*}{\textbf{\begin{tabular}[c]{@{}c@{}}Imp.\\ (GA-RS)\end{tabular}}} \\ \cline{3-6}
\multicolumn{1}{|c|}{}                                          & \multicolumn{1}{c|}{}                             & \multicolumn{1}{c|}{\textbf{best}} & \multicolumn{1}{c|}{\textbf{runtime}} & \multicolumn{1}{c|}{\textbf{best}} & \multicolumn{1}{c|}{\textbf{runtime}} &                                  \\ \hline
\multicolumn{7}{|c|}{\raisebox{.5pt}{\textcircled{\raisebox{-.4pt} {\textbf{a}}}} \textit{\textbf{Tiny Img. low Q}}}                                                                                                                                                                                                                                                          \\ \hline
\multicolumn{1}{|c|}{\textit{(BP,ON,3)}}                        & \multicolumn{1}{c|}{0.05}                         & \multicolumn{1}{c|}{63.04}         & \multicolumn{1}{c|}{2.32 hrs}         & \multicolumn{1}{c|}{61.15}         & \multicolumn{1}{c|}{2.4 hrs}          & -1.9                             \\ \hline
\multicolumn{1}{|c|}{\textit{(SB,AP,3)}}                        & \multicolumn{1}{c|}{0.51}                         & \multicolumn{1}{c|}{86.23}         & \multicolumn{1}{c|}{2.33 hrs}         & \multicolumn{1}{c|}{85.4}          & \multicolumn{1}{c|}{2.3 hrs}          & -0.8                             \\ \hline
\multicolumn{1}{|c|}{\textit{(BP,ON,4)}}                        & \multicolumn{1}{c|}{0.05}                         & \multicolumn{1}{c|}{62.00}         & \multicolumn{1}{c|}{3.15 hrs}         & \multicolumn{1}{c|}{64.72}         & \multicolumn{1}{c|}{2.81 hrs}         & 2.7                              \\ \hline
\multicolumn{1}{|c|}{\textit{(SB,AP,4)}}                        & \multicolumn{1}{c|}{0.51}                         & \multicolumn{1}{c|}{84.55}         & \multicolumn{1}{c|}{3.15 hrs}         & \multicolumn{1}{c|}{86.26}         & \multicolumn{1}{c|}{2.81 hrs}         & 1.7                              \\ \hline
\multicolumn{1}{|c|}{\textit{(BP,ON,5)}}                        & \multicolumn{1}{c|}{0.05}                         & \multicolumn{1}{c|}{61.3}          & \multicolumn{1}{c|}{4.07 hrs}         & \multicolumn{1}{c|}{62.02}         & \multicolumn{1}{c|}{3.84 hrs}         & 0.7                              \\ \hline
\multicolumn{1}{|c|}{\textit{(SB,AP,5)}}                        & \multicolumn{1}{c|}{0.51}                         & \multicolumn{1}{c|}{82.02}         & \multicolumn{1}{c|}{4.11 hrs}         & \multicolumn{1}{c|}{84.63}         & \multicolumn{1}{c|}{3.79 hrs}         & 2.6                              \\ \hline
\multicolumn{7}{|c|}{\raisebox{.5pt}{\textcircled{\raisebox{-.9pt} {\textbf{b}}}} \textit{\textbf{MNIST, high Q}}}                                                                                                                                                                                                                                                            \\ \hline
\multicolumn{1}{|c|}{\textit{(4,9,6)}}                          & \multicolumn{1}{c|}{0.05}                         & \multicolumn{1}{c|}{67.71}         & \multicolumn{1}{c|}{5.06 hrs}         & \multicolumn{1}{c|}{68.97}         & \multicolumn{1}{c|}{4.53 hrs}         & 1.3                              \\ \hline
\multicolumn{1}{|c|}{\textit{(0,1,6)}}                          & \multicolumn{1}{c|}{0.52}                         & \multicolumn{1}{c|}{81.89}         & \multicolumn{1}{c|}{5.12 hrs}         & \multicolumn{1}{c|}{85.47}         & \multicolumn{1}{c|}{4.51 hrs}         & 3.6                              \\ \hline
\multicolumn{1}{|c|}{\textit{(4,9,7)}}                          & \multicolumn{1}{c|}{0.05}                         & \multicolumn{1}{c|}{65.84}         & \multicolumn{1}{c|}{7.34 hrs}         & \multicolumn{1}{c|}{66.68}         & \multicolumn{1}{c|}{6.58 hrs}         & 0.8                              \\ \hline
\multicolumn{1}{|c|}{\textit{(0,1,7)}}                          & \multicolumn{1}{c|}{0.52}                         & \multicolumn{1}{c|}{80.21}         & \multicolumn{1}{c|}{7.38 hrs}         & \multicolumn{1}{c|}{84.16}         & \multicolumn{1}{c|}{6.58 hrs}         & 4                                \\ \hline
\multicolumn{1}{|c|}{\textit{(4,9,8)}}                          & \multicolumn{1}{c|}{0.05}                         & \multicolumn{1}{c|}{69.18}         & \multicolumn{1}{c|}{7.22 hrs}         & \multicolumn{1}{c|}{68.07}         & \multicolumn{1}{c|}{6.41 hrs}         & -1.1                             \\ \hline
\multicolumn{1}{|c|}{\textit{(0,1,8)}}                          & \multicolumn{1}{c|}{0.52}                         & \multicolumn{1}{c|}{80.46}         & \multicolumn{1}{c|}{7.22 hrs}         & \multicolumn{1}{c|}{82.85}         & \multicolumn{1}{c|}{6.39 hrs}         & 2.4                              \\ \hline
\multicolumn{7}{|c|}{\raisebox{.5pt}{\textcircled{\raisebox{-.4pt} {\textbf{c}}}} \textit{\textbf{Tiny Img, high Q}}}                                                                                                                                                                                                                                                         \\ \hline
\multicolumn{1}{|c|}{\textit{(BP,ON,6)}}                        & \multicolumn{1}{c|}{0.05}                         & \multicolumn{1}{c|}{60.58}         & \multicolumn{1}{c|}{5.08 hrs}         & \multicolumn{1}{c|}{63.35}         & \multicolumn{1}{c|}{4.46 hrs}         & 2.8                              \\ \hline
\multicolumn{1}{|c|}{\textit{(SB,AP,6)}}                        & \multicolumn{1}{c|}{0.51}                         & \multicolumn{1}{c|}{81.31}         & \multicolumn{1}{c|}{5.13 hrs}         & \multicolumn{1}{c|}{84.05}         & \multicolumn{1}{c|}{4.48 hrs}         & 2.7                              \\ \hline
\multicolumn{1}{|c|}{\textit{(BP,ON,7)}}                        & \multicolumn{1}{c|}{0.05}                         & \multicolumn{1}{c|}{59.83}         & \multicolumn{1}{c|}{7.26 hrs}         & \multicolumn{1}{c|}{61.15}         & \multicolumn{1}{c|}{6.6 hrs}          & 1.3                              \\ \hline
\multicolumn{1}{|c|}{\textit{(SB,AP,7)}}                        & \multicolumn{1}{c|}{0.51}                         & \multicolumn{1}{c|}{80.09}         & \multicolumn{1}{c|}{7.36 hrs}         & \multicolumn{1}{c|}{84.67}         & \multicolumn{1}{c|}{6.61 hrs}         & 4.6                              \\ \hline
\multicolumn{1}{|c|}{\textit{(BP,ON,8)}}                        & \multicolumn{1}{c|}{0.05}                         & \multicolumn{1}{c|}{60.45}         & \multicolumn{1}{c|}{7.2 hrs}          & \multicolumn{1}{c|}{61.61}         & \multicolumn{1}{c|}{6.8 hrs}          & 1.2                              \\ \hline
\multicolumn{1}{|c|}{\textit{(SB,AP,8)}}                        & \multicolumn{1}{c|}{0.51}                         & \multicolumn{1}{c|}{82.26}         & \multicolumn{1}{c|}{7.21 hrs}         & \multicolumn{1}{c|}{85.1}          & \multicolumn{1}{c|}{6.85 hrs}         & 2.8                              \\ \hline
\end{tabular}
\end{table}

\subsection{Results and analysis}

\textbf{\textit{Low qubit count, MNIST:}} We tabulate the results for this scenario in Table \ref{tab:mnist_low_qubit_count}. Considering noiseless conditions we generally note that the best fitness score obtained from GA lies between the mean and best score of random selection, implying that GA finds a fitness score that is better than average random selection score and is generally close to the score of best performing randomly selected embedding permutation. For example for the case of binary classes $0$ and $1$ with $4$ qubit QNN ($0$,$1$,$4$), the GA finds an embedding permutation with fitness score ($98.55$) that is almost close to the score of best randomly chosen embedding permutation ($98.56$). However, when noise is taken into account random selection of embedding permutations yield poor fitness scores. The GA on the contrary, owing to its methodical evolutionary search of iteratively finding better embedding permutation in every successive generation succeeds in finding an optimum under noise that is better than the best performing randomly chosen embedding permutation. Considering once again the case ($0$,$1$,$4$), we see that the best randomly chosen embedding permutation has a fitness score of $85.75$ while GA selected embedding permutation has $90.61$. For illustrative purposes, we also show the histograms for randomly selected embedding permutations against the final GA fitness score for ($0$,$1$,$4$) case for both noiseless and noisy conditions in Fig. \ref{fig:0_1_4_results}. Additionally, we also note that this trend is consistent across all range of SC values of binary classes. Therefore, for remaining analyses, we restrict ourselves to two pairs of binary classes, one with low SC and other with high SC, and noisy conditions. We note average improvement in GA fitness score of $0.86$ and $1.1$, and average runtimes of $2.94$ hrs ($8.1$\% less) and $2.71$ hrs ($15.0$\% less) for $4-9$ and $0-1$ classes (noisy scenario) respectively for GA, compared to $3.2$ hrs and $3.19$ hrs respectively for random selection.

\textbf{\textit{Low qubit count, Tiny Imagenet (Table \ref{tab:three_scenario_table}(a)):}} We calculate the GA fitness scores for low qubit count QNNs trained on Tiny Imagenet dataset.  From $3-5$ qubits, we note average GA fitness score of $62.63$ and $85.43$ for bell paper-orange and sulphur butterfly-alp binary classes respectively. Compared to random selection, we note an average improvement in GA fitness score of $0.5$ and $1.16$  for SC values of $0.05$ and $0.51$, respectively. Addtionally, we note an average runtime of $3.01$ hrs ($5.3$\% less) and $2.96$ hrs ($7.21$\% less) for bell paper-orange and sulphur butterfly-alp binary classes respectively for GA, compared to $3.18$ hrs and $3.19$ hrs respectively for random selection.

\textbf{\textit{High qubit count, MNIST (Table \ref{tab:three_scenario_table}(b)):}} From $6-8$ qubits we note average GA fitness score of $67.9$ and $84.16$ for $4-9$ and $0-1$ binary classes respectively. Considering these classes from Table \ref{tab:mnist_low_qubit_count}, we note average fitness scores of $69.16$ and $87.9$ respectively. This implies a decrease of $1.26$ for $4-9$ binary class while increase of $3.74$ for $0-1$ binary class. Overall we observe no drastic change in high qubit count results compared to low qubit count results. Compared to random selection, we note an average improvement in GA fitness score of $0.33$ and $3.33$ for SC values of $0.05$ and $0.52$ respectively. We note average runtime of $5.84$ hrs ($10.7$\% less) and $5.82$ hrs ($11.4$\% less) for $4-9$ and $0-1$ classes respectively for GA, as compared to $6.54$ hrs and $6.57$ hrs respectively for random selection. 
%Comparing these runtimes with low qubit count equivalent runs, we note average runtimes of $2.94$ hrs and $2.71$ hrs for $4-9$ and $0-1$ classes respectively for GA, and $3.2$ hrs and $3.19$ hrs respectively for random selection.

\textbf{\textit{High qubit count, Tiny Imagenet (Table \ref{tab:three_scenario_table}(c)):}} From $6-8$ qubits, we note average GA fitness score of $62.03$ and $84.60$ for bell paper-orange and sulphur butterfly-alp binary classes respectively. Compared to the low qubit count case, these values are a slight decrease in the average fitness score ($0.6$ and $0.83$ respectively). Once again, similar to MNIST dataset we observe no drastic change in high qubit count results compared to low qubit count results. Compared to random selection, we note an average improvement in GA fitness score of $1.76$ and $3.36$ for SC values of $0.05$ and $0.51$ respectively. We note average runtime of $5.95$ hrs ($8.6$\% less) and $5.98$ hrs ($8.8$\% less) respectively for bell paper-orange and sulphur butterfly-alp binary classes respectively for GA, compared to $6.51$ hrs and $6.56$ hrs respectively for random selection.

\subsection{Comparison with existing works}
We compare the proposed GA with Quantum Embedding Kernel (QEK) \cite{schuld2021supervised}, Quantum Approximate Optimization Algorithm (QAOA)-based embedding \cite{lloyd2020quantumembeddingsmachinelearning}, and trainable QRAC \cite{qrac}, in Table \ref{tab:comparison_table}. The comparison is done by training binary classes $3$ and $6$ of MNIST dataset under noiseless conditions\footnote{Due to implementation challenges of these works under noisy conditions and lack of results in the original work, we restrict the comparison to noisless conditions only.} for $4$ qubit-QNN, where perform training for proposed GA and QAOA-embedding-based methods while we directly pick results from the QRAC work. We note that the proposed GA performs the best (score $=$ $97$), followed by QEK (score $=$ $96.68$), then QAOA-based embedding (score $=$ $93.69$) and finally QRAC (score $=$ $90.75$). Overall, GA provides an improvement 1.003X over QEK, 1.03X over QAOA-embedding and 1.06X over QRAC embedding. This order of performance makes sense, primarily owing to the number of features being encoded and the way the features are being encoded. For proposed GA, we encode $8$ features on $4$ qubits, and QEK uses a quantum kernel to make the data linearly separable, making it easier to classify using support vector machine. On the other hand, QAOA embedding can embed only upto $3$ features, implying that less information of input data is being embedded into quantum states, yielding higher performance for GA-based algorithm. QRAC, instead of directly embedding data like the other methods, represents the input data in the form of trainable parameters which trains along with the PQC to create more accurate representation of the input data in quantum Hilbert space. Additionally, the input requires binary strings, implying that if data is decimal, it has to be rounded of to $0$s and $1$s, further introducing additional errors. 

\begin{table}[t]
\centering
\caption{Comparison of proposed GA with existing algorithms for MNIST 3,6 binary class dataset.}
\label{tab:comparison_table}
\begin{tabular}{|c|c|c|c|c|}
\hline
\textbf{Algorithm} & \textbf{\begin{tabular}[c]{@{}c@{}}Train\\ (A\%)\end{tabular}} & \textbf{\begin{tabular}[c]{@{}c@{}}Test\\ (B\%)\end{tabular}} & \textbf{\begin{tabular}[c]{@{}c@{}}Combined Score\\ =(A+B)/2\end{tabular}} & \textbf{\begin{tabular}[c]{@{}c@{}}GA \\ Improvement\end{tabular}} \\ \hline
\textit{\textbf{GA (proposed)}}        & \textbf{98.0}                                                           & \textbf{96.0}                                                          & \textbf{97.0}                                                                       & \textbf{1X}                                                                 \\ \hline
\textit{QEK} \cite{schuld2021supervised}    & 97.37                                                          & 96.0                                                          & 96.68                                                                      & 1.003X                                                             \\ \hline
\textit{QAOA Emb.} \cite{lloyd2020quantumembeddingsmachinelearning} & 92.88                                                          & 94.5                                                          & 93.69                                                                      & 1.03X                                                              \\ \hline
\textit{QRAC} \cite{qrac}      & 90.3                                                           & 91.2                                                          & 90.75                                                                      & 1.06X                                                              \\ \hline

\end{tabular}
\end{table}

\subsection{Limitations}
% This point may not be needed! You can check and confirm.
A potential drawback of the proposed GA method is it's overall runtime however it is comparable or sometimes better than random selection at better fitness scores. Assuming each embedding permutation takes average training time $T$, for population size $s_{pop}$ and $g$ generations, the overall runtime would be $\mathcal{O}(s_{pop}.g.T)$. While $T$ is scalable to an extent (as we show by increasing the qubit count), the algorithm runtime would greatly increase if the search space of GA is expanded by increasing either $s_{pop}$ or $g$. For fixed $s_{pop}$($=$$20$) and $g$($=$$5$), we show the runtime for both random selection method and proposed GA in Tables \ref{tab:mnist_low_qubit_count} and \ref{tab:three_scenario_table}. In general, the fastest method would be to select a single random embedding permutation and training it, however it may not yield best performance. Next would be random selection of $s_{pop} \times g$ embedding permutations and training them, which may yield better performing embedding permutation compared to choosing just one randomly. Finally, performing GA for $s_{pop}$ population size for $g$ generations yields best performance with similar runtime as random selection.

Additionally, the GA-based search outcome is valid for the chosen backend only (\textit{FakeBrisbane()} in this case). GA has to be rerun for a different specific backend e.g., \textit{FakeSherbrooke()}. This is also true for random selection. A potential solution to this issue could be to gather data of best embedding permutation selected by GA for various coupling configurations across different hardware, and build a predictive model that takes input as the noise characteristics of the coupling configuration in the data gathered and would output the corresponding best embedding permutation. This could be a subject for further exploration. %This may sound counterintuitive at first because the potential solution would try to run GA for multiple hardware which is what we would want to avoid. However, once sufficient data is gathered, for rest of the desired coupling configurations that haven't been tested out the predictive model would then make predictions solely based on the noise data rather than rerunning the entire GA.

%% file: 6_Conclusion.tex
\section{Conclusion}

%\textbf{\textit{Conclusion:}}
In this work, we presented the problem of QE of classical data as a search problem as compared to an optimization problem, and solved it using GA. With MNIST and Tiny Imagenet as our input datasets, the proposed GA with it's evolutionary search provided an optimal embedding permutation under noise that is better in performance compared to best randomly chosen embedding permutation. We also show that the proposed GA is scalable with the number of qubits.

\section*{Acknowledgements}
We acknowledge the usage of IBM Quantum along with Pennylane for performing all the experiments. All the relevent code has been added to a GitHub Repository\footnote{GitHub repository link: \url{https://github.com/KoustubhPhalak/quantum-embed-genetic/}}. This work is supported in parts by NSF (CNS-1722557, CNS-2129675,CCF-2210963,CCF-1718474,OIA-2040667, DGE-1723687, DGE-1821766, and DGE-2113839) and Intel’s gift.